\documentclass[12pt]{article}

\usepackage{sbc-template}
\usepackage{graphicx,url}
\usepackage{subcaption}
\usepackage[utf8]{inputenc}
\usepackage[brazil]{babel}
\usepackage{amsmath}
\usepackage{multirow}
\usepackage{booktabs} 
\title{Análise de Desaprendizado de Máquina em Modelos de Classificação de Imagens Médicas}

\author{Andreza M. C. Falcao\inst{1}, Filipe R. Cordeiro\inst{1}}

\address{
  Visual Computing Lab, Departamento de Computação, \\ Universidade Federal Rural de Pernambuco (UFRPE), Brasil
  \email{andreza.mcfalcao@ufrpe.br, filipe.rolim@ufrpe.br } 
}

\begin{document} 

\maketitle

\begin{abstract}
Machine unlearning aims to remove private or sensitive data from a pre-trained model while preserving the model’s robustness. Despite recent advances, this technique has not been explored in medical image classification. This work evaluates the SalUn unlearning model by conducting experiments on the PathMNIST, OrganAMNIST, and BloodMNIST datasets. We also analyze the impact of data augmentation on the quality of unlearning. Results show that SalUn achieves performance close to full retraining, indicating an efficient solution for use in medical applications.
\end{abstract}
     
\begin{resumo} 
O desaprendizado de máquina tem como objetivo remover dados privados ou sensíveis de um modelo pré-treinado, preservando a robustez do modelo. Apesar dos avanços, essa técnica não tem sido explorada em classificação de imagens médicas. Esse trabalho avalia o modelo de desaprendizagem Salun, conduzindo experimentos nas bases PathMNIST, OrganAMNIST e BloodMNIST. Também analisamos a influência do aumento de dados na qualidade do desaprendizado. Resultados mostram que o Salun obtém resultados próximos ao retreinamento completo, indicando uma solução eficiente para ser usada em aplicações médicas.

\end{resumo}

\section{Introdução}

Modelos de aprendizagem de máquina têm sido amplamente utilizados na área médica para tarefas de detecção e auxílio ao diagnóstico a partir de imagens~\cite{chan2020deep}. No entanto, esses modelos dependem de grandes volumes de dados para treinamento, os quais frequentemente contêm informações sensíveis de pacientes. Com o crescente uso de dados pessoais na medicina, surgem preocupações com a privacidade, impulsionadas por regulamentações como o Direito de Ser Esquecido (Right to be Forgotten)\cite{hoofnagle2019european}, que garantem aos usuários o direito de solicitar a remoção de seus dados pessoais de sistemas organizacionais\cite{dang2021right}. 

\begin{figure}
    \centering
    \includegraphics[width=0.8\linewidth]{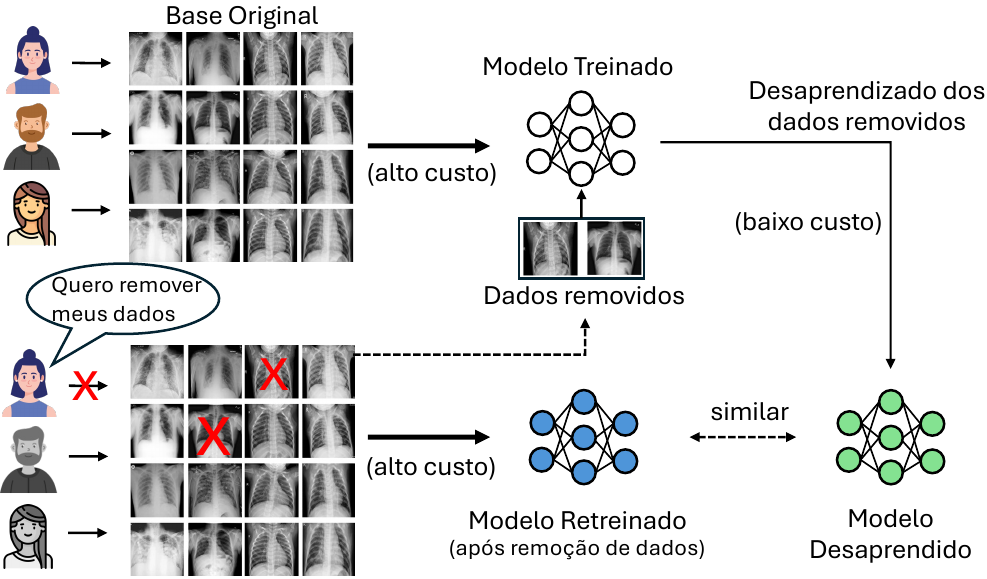}
    \caption{Processo de desaprendizado de máquina. Ao ser solicitada a remoção de dados, é gerado um modelo desaprendido, sem a influência das amostras removidas. O desaprendizado tem como objetivo gerar um modelo próximo ao de um modelo retreinado do zero, mas com um baixo custo computacional.}
    \label{fig:mu}
\end{figure}

Tradicionalmente, a forma de lidar com essa questão envolve o retreinamento completo do modelo, excluindo os dados solicitados. Embora esse método garanta conformidade com as regulamentações, ele é extremamente custoso em termos de tempo de processamento e recursos computacionais.
A área de Desaprendizado de Máquina (DM) investiga métodos para remover seletivamente a influência de dados específicos de modelos já treinados, sem a necessidade de retreinamento completo. 
A Figura~\ref{fig:mu} ilustra o processo de desaprendizado, onde, uma vez que um usuário solicita a remoção de um dado pessoal, o processo de desaprendizado de máquina busca remover a influência daqueles dados, para que se comporte como um modelo retreinado sem os dados, mas sem o alto custo computacional.

Vários trabalhos têm sido propostos a fim de investigar a eficiência de métodos de DM em tarefas de classificação de propósito geral (objetos, animais, veículos, etc)~\cite{zhang2023review}. No entanto, esse tipo de técnica foi pouco explorado para tarefas na área médica e sabe-se que modelos de DM enfrentam desafios significativos ao lidar com problemas complexos.

Neste trabalho, investigamos a viabilidade do uso de métodos de DM como uma alternativa eficiente ao retreinamento completo na remoção de dados em modelos de classificação de imagens médicas.
Para isso, nos baseamos no modelo Salun~\cite{fan2024salun}, que é um modelo de DM do estado da arte, aplicado para desaprender dados de um modelo de classificação ResNet18, utilizando as bases PathMNIST, OrganMNIST e BloodMNIST~\cite{medmnist}. Além disso, investigamos a influência do uso de aumento de dados no desaprendizado, que também é algo investigado na literatura. 




\section{Trabalhos Relacionados} \label{sec:firstpage}



Trabalhos anteriores buscam reduzir a influência dos dados removidos através de mudança aleatória de anotação~\cite{golatkar2020eternal} ou gradiente ascendente~\cite{graves2021amnesiac} O modelo Saliency Unlearning (SalUn)~\cite{fan2024salun} é um modelo de DM aproximado do estado da arte, que utiliza mapas de saliência para ajustar seletivamente pesos do modelo, baseado na ativação das amostras a serem removidas. 
Outros trabalhos buscam otimizar o processo de desaprendizado. Jia \textit{et. al }\cite{jia2023model} mostram que o processo de desaprendizado é mais eficiente ao realizar podas na rede, tornando a rede com menor número de parâmetros. O trabalho de Di \textit{et. al}~\cite{di2024label} mostra que utilizar \textit{soft labels} (anotações de entre 0 e 1) ajuda o modelo de desaprendizagem comparado ao tradicional \textit{hard label} (anotações com 0s e 1s). Neste trabalho, fazemos uma investigação também do impacto do uso de aumento de dados no processo de desaprendizado.




Apesar dos avanços, técnicas de DM ainda não foram avaliadas para bases de imagens médicas, que apresentam maior complexidade. Este trabalho busca preencher essa lacuna, em modelos treinados sobre MedMNIST, avaliando também a influência do uso de aumento de dados no processo de desaprendizagem. 


\section{Metodologia}


\subsection{Base de dados}


Utilizamos as bases de imagens PathMNIST, OrganAMNIST e BloodMNIST do repositório MedMNIST~\cite{medmnist}. O MedMNIST é uma coleção de bases de dados de imagens médicas pré-processadas e em formato padronizado, projetada para facilitar a pesquisa em aprendizado de máquina em aplicações médicas.

A base PathMNIST contém imagens de biópsias de tecidos patológicos, coloridas, organizadas em 9 classes, que representam diferentes tipos de tecidos ou estados patológicos. A base contém 107.180 imagens, divididas em 89.996 imagens de treino, 10.004 de validação e 7.180 de teste. 

A base OrganAMNIST contém imagens de seções axiais de ressonâncias magnéticas (MRI) de órgãos abdominais. A base é composta de 11 classes correspondendo a diferentes órgãos abdominais. A base é composta por 58.830 imagens, divididas em 34.561 imagens de treino, 6.491  de validação e 17.778 de teste.

A base BloodMNIST contém imagens microscópicas de células sanguíneas, com o objetivo de classificar diferentes tipos de células, auxiliando no diagnóstico de doenças hematológicas. A base possui 8 classes e 17.092 imagens, divididas em 11.959 imagens de treino, 1.712 de validação e 3.421 de teste.


Todas as imagens das bases foram redimensionadas para tamanho $64 \times 64$ \textit{pixels}. 

\subsection{Treinamento e Desaprendizado}

O modelo utilizado para a tarefa de classificação de imagens nas bases do MEDMNIST foi uma rede RestNet-18~\cite{wu2019wider}. Inicialmente, o modelo é treinado com a base inteira , por 200 épocas, utilizando taxa de aprendizagem 0.1, tamanho de \textit{batch} 256. Utilizamos a configuração padrão de aumento de dados utilizada no Salun, que é o \textit{random crop} e \textit{horizontal flip}~\cite{mumuni2022data}.  Após as 200 épocas, o modelo treinado $\theta_i$ é utilizado como base para a estratégia de desaprendizado.

A partir do modelo treinado com a base inteira $\mathcal{D_i}$, realizamos a operação de esquecimento de dados, que consiste em remover as amostras selecionadas de $\mathcal{D_i}$, resultando na base restante $\mathcal{D}_r=\mathcal{D}_i-\mathcal{D}_f$, onde $\mathcal{D}_f$ consiste no conjunto de amostras a serem esquecidas. As amostras a serem esquecidas são selecionadas utilizando uma taxa de esquecimento $\delta_f$, realizando uma seleção de forma aleatória. Essa abordagem é a mesma utilizada em \cite{fan2024salun}.

Utilizamos o método Salun para remover a influência das amostras em $\mathcal{D}_f$ em $\theta_i$, obtendo o modelo desaprendido $\theta_u= \text{Salun}(\mathcal{D}_f, \mathcal{D}_r, \theta_i)$.
O SalUn identifica as partes mais influentes do modelo para os dados a serem esquecidos e as modifica para reduzir essa influência, permitindo que o modelo esqueça os dados indesejados de forma mais eficiente e com menor impacto em seu desempenho geral. Essa identificação é feita através da análise da saliência dos pesos do modelo em relação aos dados a serem esquecidos. 
O processo de desaprendizado é feito por 10 épocas, conforme utilizado em \cite{fan2024salun}.

 Também comparamos o Salun com o retreino completo. No retreinamento completo, é treinada uma nova ResNet18 por 200 épocas, utilizando a base $\mathcal{D}_r$, sem os dados esquecidos. Os resultados do retreino são usados como base (padrão ouro) para avaliar o modelo de desaprendizado. 
 


\subsection{Métricas de Avaliação}

A métricas de avaliação de desaprendizado de máquina consistem em comparar os resultados do modelo desaprendido com o modelo retreinado. Nesse trabalho, utilizamos as métricas UA, RA, TA, MIA, AG e RTE, as quais são utilizadas em ~\cite{fan2024salun, di2024label}. Cada métrica é descrita a seguir:

\begin{itemize}
\item \textbf{Unlearning Accuracy (UA):} Mede a acurácia do modelo desaprendido nos dados a serem esquecidos. Um valor baixo de UA indica que o modelo conseguiu esquecer os dados indesejados. 

\item \textbf{Remaining Accuracy (RA):} Mede a acurácia do modelo desaprendido nos dados restantes (os dados que não foram esquecidos). Um valor alto de RA indica que o modelo conseguiu manter seu desempenho nos dados relevantes. 

\item \textbf{Testing Accuracy (TA):} Mede a acurácia do modelo desaprendido em um conjunto de teste independente. Um valor alto de TA indica que o modelo conseguiu generalizar para novos dados. 

\item \textbf{Membership Inference Attack (MIA):} Mede a vulnerabilidade do modelo desaprendido a ataques de inferência de associação. Um valor baixo de MIA indica que o modelo está menos vulnerável a esses ataques.

\item \textbf{Average GAP (AG):} Mede a proximidade do modelo desaprendido com o modelo retreinado. É calculado pelo módulo da média das diferenças entre as métricas UA, RA, TA e MU do modelo desaprendido e o modelo retreinado.

\item \textbf{Run-time efficiency (RTE):} Mede o tempo de execução de cada método.

\end{itemize}

\section{Resultados}


Analisamos o desempenho do Salun para as bases BloodMNIST, OrganAMNIST e PathMNIST, considerando uma taxa de esquecimento $\delta$  dos dados de treino de 10\% e 50\%. Os resultados para $\delta=10\%$ e $\delta=50\%$ são mostrados nas Tabelas~\ref{tab:fr10} e \ref{tab:fr50}, respectivamente. Os valores em negrito mostram a diferença comparada com o método de Retreino. Os resultados mostram uma diferença média próxima de zero, representada pela métrica AG, o que indica uma proximidade de resultados do método de retreino. No entanto, o tempo de execução é muito menor utilizando o Salun, como mostra a métrica RTE. Os resultados também indicam uma maior dificuldade de desaprendizado para uma taxa de esquecimento maior e também para a base PathMNIST, que é mais complexa, como mostram os resultados da métrica TA.

\begin{table}[ht]
    \centering
    \caption{Resultados para taxa de esquecimento de 10\%, para as métricas UA, RA, TA, MIA, AG e RTE (em minutos). Resultados em negrito mostram a diferença de cada métrica comparado com o método de Retreino, que é  o padrão ouro. A métrica AG mostra a média dessas diferenças.}
    \scalebox{0.86}{
    \begin{tabular}{cc|cccccc}
    \toprule
      \multirow{2}{*}{Base} & \multirow{2}{*}{Método} & \multicolumn{5}{c}{Taxa de Esquecimento (10\%)}  \\
      \cmidrule{3-8}
       &  & UA  & RA & TA & MIA & AG & RTE \\
         \midrule
      \multirow{2}{*}{BloodMNIST} & Retreino & 0,84 \textbf{(0,00)} & 99,80 \textbf{(0,00)} & 98,57 \textbf{(0,00)} & 1,76 (\textbf{0,00)} & \textbf{0,00} & 22,2\\
       & Salun & 0,00 \textbf{(0,84)} & 99,92 \textbf{(0,12)}  & 98,89 \textbf{(0,32)} & 0,17 \textbf{(1,59)} & \textbf{0,72} &  1,1\\
       \midrule 
       \multirow{2}{*}{OrganAMNIST} & Retreino & 0,06 \textbf{(0,00)}  & 100,00 \textbf{(0,00)} & 96,37 \textbf{(0,00)}  & 1,53 \textbf{(0,00)} & \textbf{0,00} & 63,3 \\
       & Salun & 0,00 \textbf{(0,06)} & 100,00 \textbf{(0,00)} & 95,13 \textbf{(1,24)}  & 0,69 \textbf{(0,84)}&  \textbf{0,53} & 2,8 \\
       \midrule 
       \multirow{2}{*}{PathMNIST} & Retreino & 0,11 \textbf{(0,00)} & 100,00 \textbf{(0,00)}  & 87,77 \textbf{(0,00)} & 1,06 \textbf{(0,00)} & \textbf{0,00} & 160 \\
       & Salun & 1,09 \textbf{(0,98)} & 98,84 \textbf{(1,16)} & 77,49 \textbf{(10,28)} & 4,43 \textbf{(3,37)} & \textbf{3,95} & 7,6\\
        \bottomrule
    \end{tabular}
    }
    \label{tab:fr10}
\end{table}

\begin{table}[ht]
    \centering
    \caption{Resultados para taxa de esquecimento de 50\%, para as métricas UA, RA, TA, MIA, AG e RTE (em minutos). Resultados em negrito mostram a diferença de cada métrica comparado com o método de Retreino, que é  o padrão ouro. A métrica AG mostra a média dessas diferenças.}
    \scalebox{0.86}{
    \begin{tabular}{cc|cccccc}
    \toprule
      \multirow{2}{*}{Base} & \multirow{2}{*}{Método} & \multicolumn{5}{c}{Taxa de Esquecimento (50\%)}  \\
      \cmidrule{3-8} 
       &  & UA  & RA & TA & MIA & AG& RTE \\
         \midrule
      \multirow{2}{*}{BloodMNIST} & Retreino & 1,37 \textbf{(0,00)} & 100,00 \textbf{(0,00)} & 98,48 \textbf{(0,00)}& 3,61 \textbf{(0,00)} & \textbf{0,00}& 22,2\\
       & Salun &  0,12 \textbf{(1,25)} & 99,93 \textbf{(0,07)} &98,77 \textbf{(0,29)} & 0,45 \textbf{(3,16) }& \textbf{1,57} &  1,3\\
       \midrule 
       \multirow{2}{*}{OrganAMNIST} & Retreino  & 0,13 \textbf{(0,00)} & 100,00 \textbf{(0,00)}  & 95,93 \textbf{(0,00)} & 1,93 \textbf{(0,00)} & \textbf{0,00} & 63,3\\
       & Salun & 0,02 \textbf{(0,11)} & 99,94 \textbf{(0,06)} & 94,99 \textbf{(0,94)} & 0,89 \textbf{(1,04)} & \textbf{0,70}& 3,5\\
       \midrule 
       \multirow{2}{*}{PathMNIST} & Retreino &  0,20 \textbf{(0,00)} & 100,00 \textbf{(0,00)} & 91,80 \textbf{(0,00)} & 1,93 \textbf{(0,00)} & \textbf{0,00 } & 160\\
       & Salun &  2,33 \textbf{(2,13) }& 97,82 \textbf{(2,18)} & 83,87 \textbf{(7,93)} & 6,60 \textbf{(4,67)} & \textbf{4,91}& 8,9\\
        \bottomrule
    \end{tabular}
    }
    \label{tab:fr50}
\end{table}

\begin{figure} [!ht]
    \centering
    \includegraphics[width=0.8\linewidth]{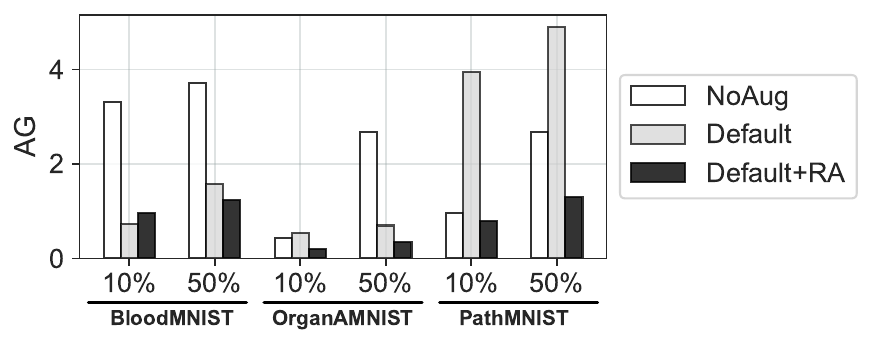}
    \caption{Resultados do método Salun, observando a métrica AG, usando 3 cenários de aumento de dados: NoAug, Default e Default+RA, considerando níveis de esquecimento de 10\% e 50\%. }
    \label{fig:aug}
\end{figure}

Analisamos também o impacto de usar \textit{data augmentation} durante as etapas de treinamento e desaprendizagem. Para isso, analisamos 3 cenários de uso de \textit{data augmentation}: 1) \textit{NoAug} (sem augmentation), 2) \textit{Default}: \textit{random crop} + horizontal flip, 3)  e 3) Default + RA: \textit{random crop} + horizontal flip + RandomAug~\cite{mumuni2022data}. A configuração \textit{default} é a configuração do Salun e a usada na maioria dos trabalhos da literatura. A combinação com RandomAug mostrou melhoria para a maioria das análises, conforme mostra a Figura~\ref{fig:aug}.

~\section{Conclusão}

O método SalUn obteve um desempenho comparável ao do Retrain nas bases analisadas: PathMNIST, OrganMNIST e BloodMNIST. As métricas analisadas indicam que o SalUn é eficaz em remover a influência de dados específicos sem comprometer significativamente o desempenho geral do modelo. Também concluímos que o uso de aumento de dados pode auxiliar na eficiência do desaprendizado de máquina e que esse modelo pode ser usado com eficiência em bases de dados médicas. 

\bibliographystyle{sbc}
\bibliography{sbc-template}

\begin{thebibliography}{}

\bibitem[Chan et~al. 2020]{chan2020deep}
Chan, H.-P., Samala, R.~K., Hadjiiski, L.~M., and Zhou, C. (2020).
\newblock Deep learning in medical image analysis.
\newblock {\em Deep learning in medical image analysis: challenges and applications}, pages 3--21.

\bibitem[Dang 2021]{dang2021right}
Dang, Q.-V. (2021).
\newblock Right to be forgotten in the age of machine learning.
\newblock In {\em Advances in Digital Science: ICADS 2021}, pages 403--411. Springer.

\bibitem[Di et~al. 2024]{di2024label}
Di, Z., Zhu, Z., Jia, J., Liu, J., Takhirov, Z., Jiang, B., Yao, Y., Liu, S., and Liu, Y. (2024).
\newblock Label smoothing improves machine unlearning.
\newblock {\em arXiv preprint arXiv:2406.07698}.

\bibitem[Fan et~al. 2024]{fan2024salun}
Fan, C., Liu, J., Zhang, Y., Wong, E., Wei, D., and Liu, S. (2024).
\newblock Salun: Empowering machine unlearning via gradient-based weight saliency in both image classification and generation.
\newblock In {\em International Conference on Learning Representations (ICLR)}.

\bibitem[Golatkar et~al. 2020]{golatkar2020eternal}
Golatkar, A., Achille, A., and Soatto, S. (2020).
\newblock Eternal sunshine of the spotless net: Selective forgetting in deep networks.
\newblock In {\em Proceedings of the IEEE/CVF conference on computer vision and pattern recognition}, pages 9304--9312.

\bibitem[Graves et~al. 2021]{graves2021amnesiac}
Graves, L., Nagisetty, V., and Ganesh, V. (2021).
\newblock Amnesiac machine learning.
\newblock In {\em Proceedings of the AAAI Conference on Artificial Intelligence}, volume~35, pages 11516--11524.

\bibitem[Hoofnagle et~al. 2019]{hoofnagle2019european}
Hoofnagle, C.~J., Van Der~Sloot, B., and Borgesius, F.~Z. (2019).
\newblock The european union general data protection regulation: what it is and what it means.
\newblock {\em Information \& Communications Technology Law}, 28(1):65--98.

\bibitem[Jia et~al. 2023]{jia2023model}
Jia, J., Liu, J., Ram, P., Yao, Y., Liu, G., Liu, Y., Sharma, P., and Liu, S. (2023).
\newblock Model sparsity can simplify machine unlearning.
\newblock {\em Advances in Neural Information Processing Systems}, 36:51584--51605.

\bibitem[Mumuni and Mumuni 2022]{mumuni2022data}
Mumuni, A. and Mumuni, F. (2022).
\newblock Data augmentation: A comprehensive survey of modern approaches.
\newblock {\em Array}, 16:100258.

\bibitem[Wu et~al. 2019]{wu2019wider}
Wu, Z., Shen, C., and Van Den~Hengel, A. (2019).
\newblock Wider or deeper: Revisiting the resnet model for visual recognition.
\newblock {\em Pattern recognition}, 90:119--133.

\bibitem[Yang et~al. 2021]{medmnist}
Yang, J., Shi, R., Wei, D., Liu, Z., Wang, L., Zhou, Y., Zhou, S., Bian, C., Li, L., Wang, X., et~al. (2021).
\newblock Medmnist: A lightweight automl benchmark for medical image analysis.
\newblock \url{https://medmnist.com}.
\newblock Accessed: February 13, 2025.

\bibitem[Zhang et~al. 2023]{zhang2023review}
Zhang, H., Nakamura, T., Isohara, T., and Sakurai, K. (2023).
\newblock A review on machine unlearning.
\newblock {\em SN Computer Science}, 4(4):337.

\end{thebibliography}

\end{document}